\documentclass[a4paper]{article}
\usepackage{graphicx}
\usepackage{mathrsfs}
\usepackage{hyperref}

\def\beq{\begin{equation}}
\def\eeq{\end{equation}}
\DeclareSymbolFontAlphabet{\mathrsfs}{rsfs}
\newcommand{\scri}{\mathrsfs{I}}
\newcommand{\rscri}{r_{\!\!\scri}}
\newcommand{\aconf}{\bar\Omega}
\newcommand{\CZ}{Z4c}

\begin{document}

\title{Free hyperboloidal evolution in spherical symmetry}
\author{Alex Va\~n\'o-Vi\~nuales$^{*,\dagger}$ and Sascha Husa$^\dagger$
\\ \\ {\small$^*$School of Physics and Astronomy, Cardiff University, Queen's Buildings, CF24 3AA,}\\
 {\small Cardiff, United Kingdom - E-mail: Alex.Vano-Vinuales@astro.cf.ac.uk } \\ \\ {\small$^\dagger$Universitat de les Illes Balears and Institut d’Estudis Espacials de Catalunya,}\\ {\small Cra. de Valldemossa km.7.5, 07122 Palma de Mallorca, Spain}}
\date{}
\maketitle

\begin{abstract}
We address the hyperboloidal initial value problem in the context of Numerical Relativity, motivated by its evolution on hyperboloidal slices: smooth spacelike slices that reach future null infinity, the ``location'' in spacetime where radiation is to be extracted. Our approach uses the BSSN and Z4 formulations and a time-independent conformal factor. The resulting system of PDEs includes formally diverging terms at null infinity. Here we discuss a regularized numerical scheme in spherical symmetry. A critical ingredient are the gauge conditions, which control the treatment of future null infinity. Stable numerical evolutions have been performed with regular and black hole initial data on a hyperboloidal slice. A sufficiently large scalar field perturbation will create a black hole, whose final stationary state is different from the trumpet initial data derived here. 
\end{abstract}


\section{Introduction}

The energy loss and radiation of an isolated system are in general only well defined at future null infinity ($\scri^+$)\cite{PhysRevLett.10.66,Wald}, which also corresponds to the appropriate idealization of astronomical observers\cite{Barack:1998bv}. 
It is thus at $\scri^+$ where radiation signals should be ideally extracted from numerical simulations.
Following Penrose \cite{PhysRevLett.10.66,Penrose:1965am}, we set up our problem in a conformally compactified spacetime: instead of the physical metric $\tilde g_{\mu\nu}$ that diverges at infinity, we will use a rescaled metric $g_{\mu\nu}$ defined as
\begin{equation}\label{rescmetric}
g_{\mu\nu} \equiv\Omega^2\tilde g_{\mu\nu} ,
\end{equation}
where the conformal factor $\Omega$ vanishes at $\scri^+$.
The equations of motion ($G_{\mu\nu}[\tilde g] = 8\pi\,T_{\mu\nu}[\tilde g]$) expressed in terms of $g_{\mu\nu}$ will thus formally diverge at $\scri^+$:
\begin{equation}\label{eq:EEconformal}
G_{\mu\nu}[g] = 8\pi\,T_{\mu\nu}\left[\frac{g}{\Omega^2}\right] -\frac{2}{\Omega}\left(\nabla_\mu\nabla_\nu\Omega-g_{\mu\nu}\nabla^\gamma\nabla_\gamma\Omega\right)-\frac{3}{\Omega^2}g_{\mu\nu}(\nabla_\gamma\Omega)\nabla^\gamma\Omega  .
\end{equation}

A convenient way to cast the problem into an initial value formulation is to foliate spacetime along hyperboloidal surfaces. In the inner strong field zone they behave like common Cauchy slices and asymptotically they reach $\scri^+$, while remaining spacelike and smooth everywhere. This approach, the hyperboloidal initial value problem pioneered by Friedrich \cite{friedrich1983,lrr-2004-1,Friedrich:2003fq}, has proven to be difficult, in particular for hyperbolic free evolution schemes. For the use of elliptic-hyperbolic systems see \cite{Andersson:springer,Rinne:2009qx,Rinne:2013qc}.
Stable evolutions of regular initial data with flat end states, obtained with our spherically symmetric free evolution approach, have been presented in \cite{Vano-Vinuales:2014koa}, while the first black hole evolutions, including the measurement of scalar field power-law decay tails, were briefly described in \cite{Vano-Vinuales:2014ada}. A convenient way to represent black holes is by means of trumpet slices\cite{Hannam:2008sg}, whose interior part is mapped to an infinitely long cylinder. Here we show our progress in simulating black hole spacetimes by comparing constant-mean-curvature trumpet initial data to the end state of a collapse simulation.

\section{Hyperboloidal foliations}\label{s:hypfol}

\begin{figure}[h!!!!]
\center
\begin{tabular}{@{}c@{}@{}c@{}}
\mbox{\includegraphics[width=0.5\linewidth]{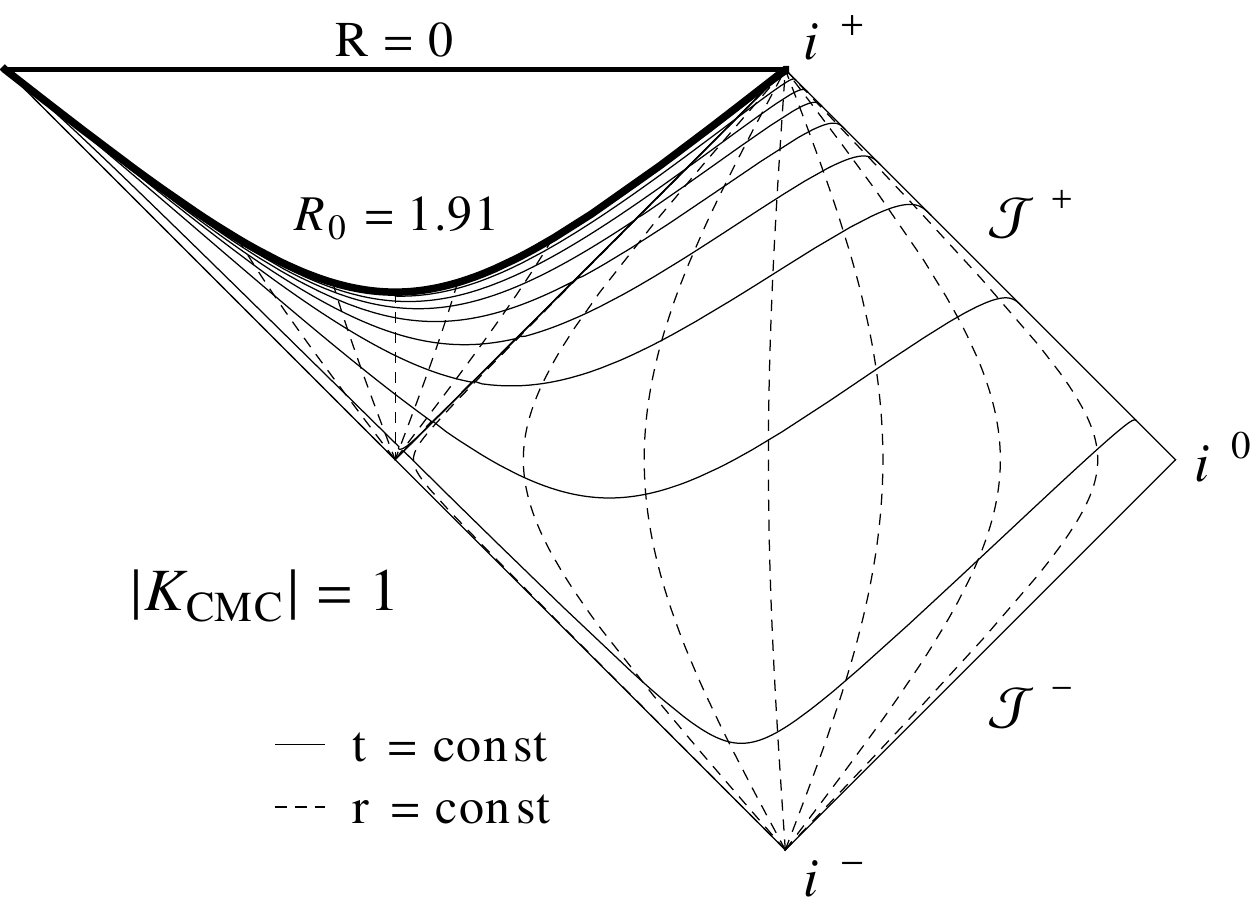}} &
\mbox{\includegraphics[width=0.5\linewidth]{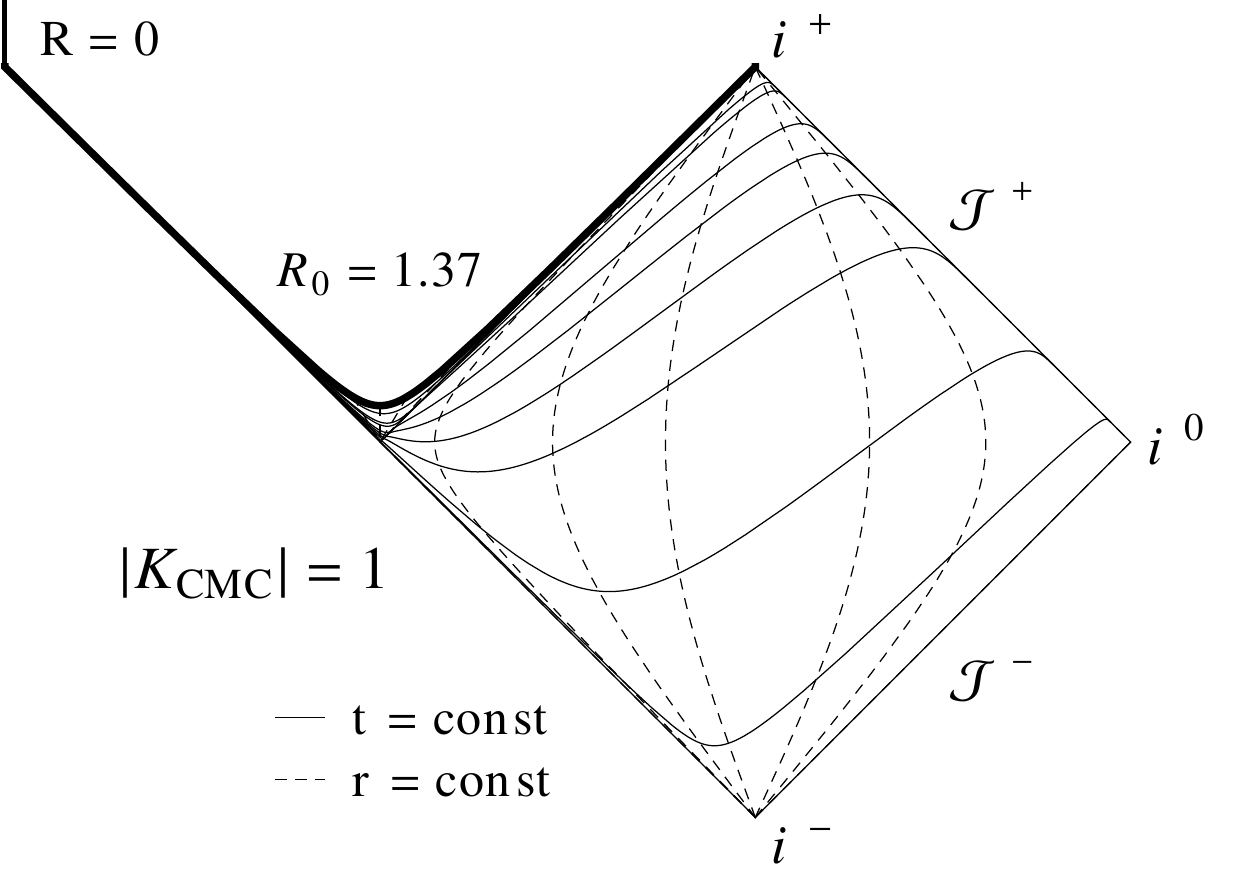}}
\end{tabular}
\caption{Penrose diagrams showing foliations for $K_{CMC}=-1$ and mass of the black hole $M=1$: on the left for Schwarzschild and on the right for Reissner-Nordstr\"om with charge $Q=0.9M$.}
\label{hypfol}
\end{figure}

We start with the following line element on a Cauchy slice:
\beq
d\tilde s^2 = -A(\tilde r)d\tilde t^2+\frac{1}{A(\tilde r)}d\tilde r^2+\tilde r^2 d\sigma^2  , \quad \textrm{where} \quad d\sigma^2\equiv d\theta^2+\sin^2\theta d\phi^2  ,
\eeq
where $A(\tilde r)$ is general enough to include a Schwarzschild or a Reissner-Nordstr\"om black hole, among others. The first step is to transform the time coordinate $\tilde t$ to a new time coordinate $t$ whose constant values determine the hyperboloidal slices with help of a height function $h(\tilde r)$ (compare e.g. \cite{Malec:2003dq}). The radial coordinate is then compactified with the factor $\aconf$ and the line element is conformally rescaled according to (\ref{rescmetric}):
\begin{equation}
t = \tilde t-h(\tilde r), \qquad  \tilde r=\frac{r}{\aconf}  \qquad\textrm{and}\qquad ds^2 = \Omega^2d\tilde s^2.
\end{equation}
The resulting line element, with $A$ and $h'$ functions of $r/\aconf$, is given by
\begin{equation}\label{fsthyp}
ds^2= -A\Omega^2dt^2+\frac{\Omega^2}{\aconf^2}\left[-2A\,h'\,(\aconf-r\,\aconf')dt\,dr+\frac{\left[1-\left(A\,h'\right)^2\right]}{A}\frac{(\aconf-r\,\aconf')^2}{\aconf^2}d r^2 + r^2 d\sigma^2\right] .
\end{equation}
We will consider a constant-mean-curvature slice, for which the spatial derivative of the height function takes the form \cite{}
\begin{equation}
h'(\tilde r)=\frac{dh}{d\tilde r}=-\frac{\frac{K_{CMC}\tilde r}{3}+\frac{C_{CMC}}{\tilde r^2}}{A(\tilde r)\sqrt{A(\tilde r)+\left(\frac{K_{CMC}\tilde r}{3}+\frac{C_{CMC}}{\tilde r^2}\right)^2}}.
\end{equation}
The constant parameter $K_{CMC}$ is the trace of the physical extrinsic curvature ($\tilde K$) and $C_{CMC}$ is an integration constant.
For a given $K_{CMC}$, there is only one possible choice for $C_{CMC}$ that will provide trumpet initial data \cite{Hannam:2006vv}. Examples of hyperboloidal trumpet foliations for the Schwarzschild and Reissner-Nordstr\"om spacetimes are shown in Fig.~\ref{hypfol}.
Following Zengino\u{g}lu \cite{Zenginoglu:2007it,Zenginoglu:2008pw}, we choose a time-independent conformal factor $\Omega$, with the following form and $\rscri$ denoting the position of null infinity:
\begin{equation}\label{omeg}
\Omega=\frac{\rscri^2-r^2}{6\, \rscri}|K_{CMC}|  .
\end{equation}
The compactification factor $\aconf$ is determined numerically by imposing a conformally flat initial spatial metric. In this way, the quantity $\aconf$ compactifies two different asymptotic ends: the trumpet at $r=0$ and the hyperboloidal foliation extending towards $\scri^+$ at $r=\rscri$.

\section{Conformally compactified equations and gauge conditions}

Both the generalized BSSN formulation \cite{PhysRevD.52.5428,Baumgarte:1998te,Brown:2007nt} and the \CZ{} equations \cite{Bernuzzi:2009ex,Weyhausen:2011cg} (a conformal version of the Z4 formulation \cite{bona-2003-67,Alic:2011gg,Sanchis-Gual:2014nha}) are used in their spherically symmetric reduction, presented in Appendix C of \cite{Vano-Vinuales:2014koa}.
A massless scalar field is coupled to the Einstein equations and, expressed in terms of the rescaled metric $g_{\mu\nu}$, its equation of motion takes the form $
g^{\mu\nu}\nabla_\mu\nabla_\nu\Phi-2\Omega^{-1}g^{\mu\nu}\nabla_\mu\Phi\nabla_\nu\Omega=0 . $

In our spherically symmetric setup we fix the location of $\scri^+$ by demanding that $\left({\partial}/{\partial t}\right)^a=\alpha n^a+\beta^a$ becomes null there. Since $\scri^+$ is null, we impose that $\left({\partial}/{\partial t}\right)^a$ is parallel to $\nabla^a\Omega$ there, and we obtain the conditions  $\left.\partial_t\Omega\right|_{\scri^+}=0$ (consistent with our time-independent $\Omega$) and $\left.-\alpha^2+\beta^a\beta_a\right|_{\scri^+}=0$. The last condition is achieved by imposing fixed values of the gauge quantities $\alpha$ and $\beta^r$ at $\scri^+$ during the evolution.

Instead of the generalized version of the Gamma-driver shift condition \cite{Alcubierre:2002kk} we used in \cite{Vano-Vinuales:2014ada}, here we implemented its integrated form\cite{vanMeter:2006vi}:
\begin{eqnarray}
\dot \beta^r = \beta^r{\beta^r}' + \lambda \, \Lambda^r - \eta \,\beta^r  + L_0 -\frac{\xi_{\beta^r}}{\Omega}\beta^r .
\end{eqnarray}
A source function $L_0$ calculated from the initial data was added, as well as a damping term ($-\beta^r/\Omega$) that makes sure that the value of $\beta^r$ stays fixed at $\scri^+$. The parameter $\lambda$ is not allowed to be larger than $(\rscri\ K_{CMC})^2/12$ if physical characteristic speeds (only outgoing modes at $\scri^+$) are desired.

In the Bona-Mass\'o family of slicing conditions\cite{Bona:1994dr}, the harmonic slicing condition \cite{}, which we used in \cite{Vano-Vinuales:2014ada}, is only marginally singularity avoiding, so that the 1+log condition (strongly singularity avoiding) deals better with the trumpet data near the origin. However, instead of the common choice $n=2$ \cite{Arbona:1999ym,Alcubierre:2001vm}, we have implemented it in the form
\begin{equation}
\dot \alpha=\beta^r\alpha'-n\,\alpha\left(\frac{\tilde K}{\Omega}-\frac{\tilde K_0}{\Omega}\right)+L_0 ,
\end{equation}
where $n= -K_{CMC}\ \rscri/3>0$, in order to obtain physical propagation speeds at $\scri^+$.
The presence of $\tilde K_0$ is necessary to avoid exponential growths (the extrinsic curvature is negative on a hyperboloidal slice) and the source function $L_0$ is calculated from initial data on the hyperboloidal foliation and ensures that the value of the lapse is fixed at $\scri^+$.

\section{Numerical tests: collapse and trumpet data}

The simulations we perform with our spherically symmetric code are restricted to Schwarzschild spacetimes ($A(\tilde r)=1-\frac{2M}{\tilde r}$) or regular spacetimes with a massless scalar field. The parameter choices are: $M=1$, $K_{CMC}=-1$, trumpet critical value of $C_{CMC}=3.11$ (see Chpt.~3 in \cite{Vano-vinuales:2015lhj}) and $\rscri=1$; for the gauge conditions: $\lambda=12^{-1}$; $\xi_\alpha,\xi_{\beta^r}=5$ for collapse and $\xi_\alpha,\xi_{\beta^r}=2$ for trumpet initial data.

\begin{figure}[h!!!!]
\center
\begin{tabular}{@{}c@{}@{}c@{}}
\mbox{\includegraphics[width=0.5\linewidth]{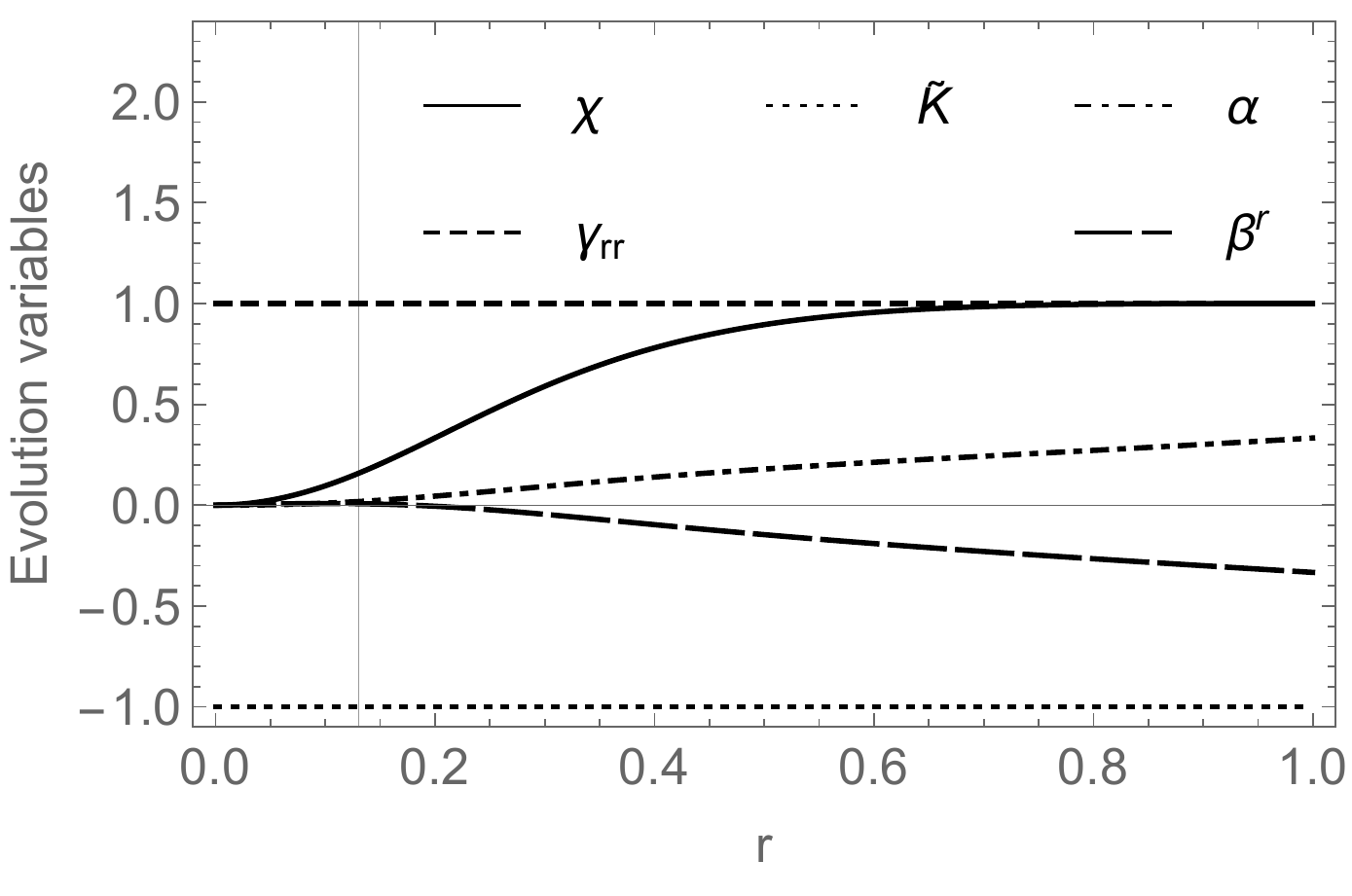}} &
\mbox{\includegraphics[width=0.5\linewidth]{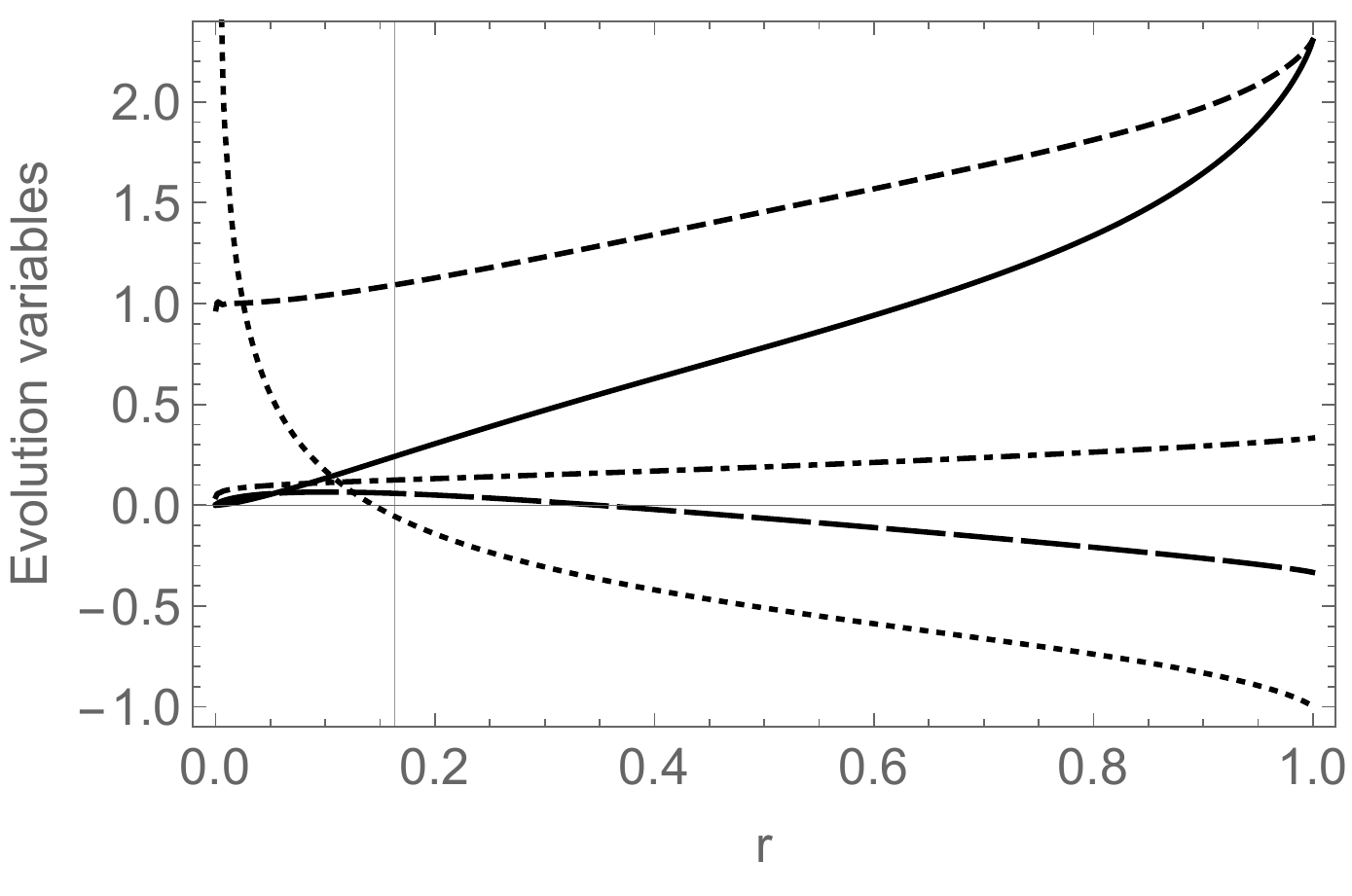}}
\end{tabular}
\vspace{-3ex}
\caption{Values of the evolved quantities for constant-mean-curvature trumpet initial data (left) and stable stationary end state (right). 
The vertical line denotes the location of the horizon.}
\label{trumpetdata}
\end{figure}

Our code uses the method of lines with a 4th order Runge-Kutta time integrator and 4th order finite differences. Kreiss-Oliger dissipation \cite{kreiss1973methods} is added to the equations, which are evolved on a staggered grid (the spatial grid avoids both the origin $r=0$ and $\scri^+$ $r=\rscri$). Outflow boundary conditions \cite{Calabrese:2005fp} are used at both the inner and outer boundaries.
In collapse simulations flat spacetime initial data is perturbed by a massless scalar field, with initial data
\vspace{-1ex}\begin{equation}
\Phi_0=B e^{-\frac{(r^2-c^2)^2}{4\sigma^4}}
\end{equation}
and chosen values for the center $c=0.5$, width $\sigma=0.1$ and amplitude $B=0.055$ in the Gaussian-like perturbation. Intermediate steps in the evolution of the collapse are shown in Fig.~8.14 in \cite{Vano-vinuales:2015lhj}. The stationary end state that is achieved coincides qualitatively with the right plot in Fig.~\ref{trumpetdata} - note how the lapse $\alpha$ goes to zero at the origin, signaling the formation of a black hole.


Vacuum constant-mean-curvature trumpet initial data as described in Sec.~\ref{s:hypfol} for some of the evolution variables are displayed on the left in Fig.~\ref{trumpetdata}. They satisfy analytically the constraint equations and, if the gauge source functions are calculated from the same trumpet initial data, the right-hand-sides of the evolution equations vanish at the analytic level. However, this initial data corresponds to an unstable solution: when evolved in our numerical implementation, the variables slowly drift away from their initial values and no stable stationary end state is found, even if the solution converges as expected. This was briefly mentioned in \cite{Vano-Vinuales:2014ada} and studied in more detail in \cite{Vano-vinuales:2015lhj}.
Nevertheless, if the gauge source functions are calculated from flat spacetime initial data, a stable stationary end state is found after some initial dynamics in the simulation (caused by the fact that the right-hand-side of the gauge equations does not vanish for the initial data). This final state coincides with the end state in the collapse simulation (provided that the mass of its final black hole is the same as $M$ in the initial trumpet data) and can be seen on the right in Fig.~\ref{trumpetdata}. This final stable stationary state also corresponds to a trumpet slice (the proper length inside of the horizon becomes infinite, while the Schwarzschild radius remains finite), but in this case the trace of the physical extrinsic curvature is not constant (see $\tilde K$ on the right plot in Fig.~\ref{trumpetdata}). More details about this will be given in \cite{bhpaper}.

\vspace{-2ex}

\section{Conclusions}

The Schwarzschild spacetime has been studied in spherical symmetry using a hyperboloidal free evolution approach based on conformally compactified versions of the generalized BSSN and \CZ{} evolutions systems for the Einstein equations. The results shown here complement those regarding the power-law decay tails of the scalar field presented in \cite{Vano-Vinuales:2014ada} and broaden our understanding about the treatment of strong field data in the hyperboloidal approach. The gauge source functions play a very important role in determining if a stable stationary solution can be found in the evolution. Here we compared analytic trumpet data with numerical collapse end data, which turned out to be trumpet data as well, but with a non-constant mean curvature. In views of this results, we plan to further study and develop the gauge conditions in the hyperboloidal framework and try to understand the possible stable stationary numerical representations of the Schwarzschild spacetime that we can obtain in our implementation.

\vspace{-2ex}

\section*{Acknowledgments}

AV was supported by AP2010-1697 and partially by the European Research Council Consolidator Grant 647839. AV and SH were supported by Spanish MINECO grants FPA2010-16495, FPA2013-41042-P and CSD2009-00064, European Union FEDER funds, and the Conselleria d'Economia i Competitivitat (Govern de les Illes Balears).

\vspace{-2ex}

\bibliographystyle{iopart-num_no_url} 
\bibliography{../hypcomp}

\end{document}